
\input phyzzx

\overfullrule=0pt

\def\p{\partial}
\def\wt{\widetilde}

\def\define#1#2\par{\def#1{\Ref#1{#2}\edef#1{\noexpand\refmark{#1}}}}
\def\con#1#2\noc{\let\?=\Ref\let\<=\refmark\let\Ref=\REFS
         \let\refmark=\undefined#1\let\Ref=\REFSCON#2
         \let\Ref=\?\let\refmark=\<\refsend}

 \let\refmark=\NPrefmark

\define\POLYAKOV
A. M. Polyakov, Mod. Phys. Lett. {\bf A3} (1988) 325.

\define\FUSTERO
Xavier Fustero, Rodolfo Gambini and Antoni Trias, Phys. Rev. Lett. {\bf 62}
(1989) 1964.

\define\BLAU
M. Blau and G. Thompson, Ann. Phys. (NY) {\bf 205} (1991) 130.

\define\ROCEK
T. H. Hansson, A. Karlhede and M. Ro$\check{c}$ek, Phys. Lett. {\bf B225}
(1989) 92.

\define\GRUNBERG
J. Grunberg, T. H. Hansson, A. Karlhede and U. Lindstrom, Phys. Lett. {\bf
B218} (1989) 321.

\define\DNA
M. D. Frank-Kamenetskij and A. V. Vologodskij, Sov. Phys. Usp. {\bf 24}
(1981) 679.

\define\FRADKIN
E. S. Fradkin and A. A. Tseytlin, Phys. Lett. {\bf B158} (1985) 316;
E. S. Fradkin and A. A. Tseytlin, Phys. Lett. {\bf B160} (1985) 69.
 
\define\LEE
J. H. Cho and J. J. Lee in preparation.

\title{ON WILSON SURFACES IN TOPOLOGICAL CHERN-SIMONS INTERACTION
IN (3+1) DIMENSIONS}

\author{JUNGJAI ~ LEE \foot{e-mail address: jjlee@road.daejin.ac.kr}}

\foot{ This work was supported by the Dae-Jin University Research Grants
       in 1996.}

\address{Department of Physics, Daejin University,
Mt.11-1, Seondan, Pocheon, Kyeonggi, Korea}

\nextline
\nextline
\nextline
\nextline
\nextline
\nextline

\abstract

We comment  on the regularization of  the expectation values of  Wilson
surfaces for the
bosonic string in  the (3+1) dimensions. We analyze the  singular behaviors
of propagator for
the Chern-Simons action with the additional higher order terms.

\vfill\eject

In paper \POLYAKOV,  Polyakov explicitly showed that the charged  particles
reverse their statistics
in  the (2+1)  dimensional  abelian gauge  theory  with the  topological
Chern-Simons term. 
Polyakov's spin factor, which is given as the expectation values of Wilson
line operator, plays 
a central role in this  phenomenon. After this work, Xavier Fustero and et'
al\FUSTERO\ showed  that
the  naive extension  of Polyakov's  construction to  the (3+1)  dimensional
case  led to  the 
transmutation between bosonic and fermionic one-dimensional structures.
They established this fact through the argument that the Dirac algebra of
$e_\mu$ fields may be represented in terms of a one-parameter family of Pauli
matrices, i.e. the spin chain operator $\sigma (t) $, which provides
the needed ingredient to reproduce the propagator for the fermionic string.

In the (2+1) dimensional Chern-Simons theory we need to introduce a framing
prescription
for the expectation values of Wilson line  operators\ROCEK. This is due to
the fact that the metric
is engaged in  the gauge fixing procedure  while those operators are
topological  in character. 
However, in the (3+1)dimensional case, the extension is a sort of $BF$
theory. According to the
illustration in Ref. \BLAU, we have the the antisymmetric $B$ field in
addition to the connection field
$A$,  the question  of framing  does  not arise  in the  analogous
calculations. The  result is,
therefore, finite and unambiguous.

In this  paper, we  will show that  if we  try to  calculate the expectation
values of  the
Wilson surface operator  in (3+1) dimensional bosonic  string theory the
careful  regularization 
must be needed. To do this, we will give the two definitions for the
expectation values of the
Wilson surface  operator and  discuss their  singular behaviors  near the
boundary of the string 
world sheets. Next the  effect of the higher order terms(kinetic terms) in
the action will be
analyzed. We will also investigate the singular  behaviors of spin factors
in the time like
gauge, which is independent of the metric, and compare with that in the
covariant gauge. 
Finally we will discuss a problem with the gauge invariance of the antisymmetric
2-form field $B_{\mu \nu}$.

According to Ref. \FUSTERO, the spin factor $\Phi (S,C)$ between  the
initial and final spatial
configurations $P_i$ and $P_f$ of the string is given as
$$\eqalign
{
  &exp(i \Phi (S,C)) \cr
  &= < exp(i \int_S d \sigma^{\mu \nu} B_{\mu \nu})
  exp(i \oint_C dx^\mu A_\mu)>   \cr
  &= \int DA DB exp( -i \int d^4 x
  \epsilon^{\mu \nu \rho \sigma} B_{\mu \nu} \p_\rho A_\sigma )
  exp(i \int_S d \sigma^{\mu \nu} B_{\mu \nu} )
  exp(i \oint_C dx^{\mu} A_{\mu} )
}
 \eqn \eone
$$
Here the $S$ and $C$ in $\Phi (S,C)$ denote  the world sheets and its  border
respectively, and we regard the spin factor $\Phi (S,C)$ as the expectation
value of the Wilson surface operator in the $BF$ theory.

In above path integral \eone, we considered a (3+1)dimensional bosonic-string
classical system in the  electromagnetic interactions  described by  the gauge
potential $ A_\mu (x) $ and the antisymmetric background field $B_{\mu
\nu}(x)$.  These  fields  are coupled  through  the  topological
Chern-Simons term
$$
  \int d^4 x \epsilon^{\mu \nu \rho \sigma} B_{\mu \nu} \p_\rho A_\sigma
  \eqn\etwo
$$
which is invariant under the gauge transformations with the gauge parameters
$\Lambda$ and $\xi_\mu$;
$$
  \delta A_\mu  = \p_\mu \Lambda,
  \eqn\ethree
$$
$$
  \delta B_{\mu \nu} = \p_\mu \xi_\nu - \p_\nu \xi_\mu .
  \eqn\efour
$$
The integrals \eone\ over the $A_\mu$ and $B_{\mu \nu}$ fields lead to the spin
factor
$$\eqalign
  {
   \Phi (S,C) =& - {i \over 4} \int_{S_C} d\sigma_{\mu \nu} (y)
   \int_{ {S'}_C} d\sigma_{\mu \nu} (y') \delta (y-y') \cr
   =& - {i \over {4\pi^2} } \oint_C dy_\mu
   \int_{{S'}_C} d\sigma_{\lambda \rho}(y')
   \epsilon_{\mu \nu \lambda \rho} \p_\nu {1 \over {|y-y'|^2} }
  }
  \eqn\efive
$$
where $ {S'}_C $ is the second arbitrary surface with border $C$.

This expression for the spin factor \efive\ equals to the  writhe of ribbon
as the
string world sheet\GRUNBERG\DNA. The writhe depends on the metric, this
dependence
enters the propagator via the gauge fixing conditions
$\p_\mu A^\mu = 0 $ and $\p_\mu B^{\mu \nu} = 0 $.
The phase factor $\Phi (S,C)$  is singular, since  $C$ must tend to the
border of
$\sigma$, therefore the careful  regularization process  is needed.
First we introduce the framing
$$\eqalign
  {
   \wt{\Phi} (S,C)
   \equiv & \lim_{\epsilon\to 0} (
     - {i \over {4\pi^2} } \oint_C dy_\mu
     \int_{ {S'}_C} d\sigma_{\lambda \rho}(y')
     \epsilon_{\mu \nu \lambda \rho} \p_\nu
     {1 \over {|y-y'-\epsilon \hat{n} (y)|^2} }
    ) \cr
   =& L(S,C, \hat{n} ) ,
  }
  \eqn\esix
$$
where $\hat{n} =( n_1,  n_2, e, f )$ ; $n_{1\mu}$ and $n_{2\mu}$ are two normal
vectors orthogonal to the both $e_\mu$ and $f_\mu$ which are the unit tangent
vectors to the closed path $C$ in (3+1) dimensions. The $\hat{n}$ defines a
framing of the surface $S$ with its border $C$. In the equation \esix, $L$
is the
linking number, a topological quantity, which measure the  revolution times
of the closed path $C$ around the sheet $S$ in four dimensions.

Let us compare  the two expressions for the  spin factor. The expression \efive\
for the spin factor is  metric dependent and  framing independent, while on
the other the expression \esix\ framing dependent and metric independent, so we
can ask which  definition is correct. This metric dependence in eq. (4)
originates
in the short distance singularities of  the propagator which enters via the
gauge fixing condition at the quantum level.

To clarify the singular behavior of the propagator, we add the higher order
terms to the action \etwo. This work is physically meaningful as higher
order terms
are metric dependent but frame independent. So we take the following action
for gauge fields from the  effective field theory of quantized strings\FRADKIN:
$$
 \Gamma =\int d^4 x ( \alpha_1 F^{\mu \nu} F_{\mu \nu}
   + \alpha_2 H^{\mu \nu \rho} H_{\mu \nu \rho}
   + \alpha_3 \epsilon^{\mu \nu \rho \sigma} A_\mu \p_\nu B_{\rho \sigma}
   ),
  \eqn\eseven
$$
where $\alpha_1$, $\alpha_2$ and $\alpha_3$ are constants.
Here
$$\eqalign{
  F_{\mu \nu}=& \p_\mu A_\nu - \p_\nu A_\mu , \cr
  H_{\mu \nu \rho}=& \p_\mu B_{\nu \rho} + cyclic~~ permutations~~ of
~~\mu,~ \nu,~ \rho .
  }
  \eqn\eeight
$$

In the covariant (Feynman) gauge we obtain the propagator
$$\eqalign{
  D_{\mu \nu \rho} (x,y)
  =& D_{\mu \nu \rho} (|x-y|)=< A_\mu (x) B_{\nu \rho}(y)>,  \cr
  D_{\mu \nu \rho} (r)
  =& {1 \over { 32 \pi^2 \alpha_1 \alpha_3 } } \epsilon^{\mu \nu \rho \sigma}
  \p_\sigma ( {1 \over {\mu^2 r^2}} - {1 \over {\mu r}} K_1(\mu r) )    \cr
  =& {1 \over { 32 \pi^2 \alpha_1 \alpha_3 } } \epsilon^{\mu \nu \rho \sigma}
  \hat{r}_\sigma ( {{-2} \over {\mu^2 r^3}} - {1 \over r} K_2(\mu r) ),
 }
  \eqn\enine
$$
where $r=|x-y|$, $\mu^2 = { \alpha_3^2 \over \alpha_2 }$ ; $\mu$ is a
topological
mass, $K_1 (x)$ and $K_2 (x)$ are the modified Bessel functions. Though we can
get three propagators in this theory, we will consider
only $<A_\mu B_{\nu \rho}>$ propagator  since  we  are  concerned only  with
the topological nature of Chern-Simons term, i.e. the statistics of the theory.
To investigate how the presence of the $F_{\mu \nu} F^{\mu \nu}$ and
$H_{\mu \nu \rho}H^{\mu \nu \rho}$ terms modify the fields close to the border
$C$, consider
$$\eqalign{
  A_\mu=& \int_{S_C} d\sigma_{\nu \rho} D_{\mu \nu \rho}(r)
        + Coulomb~~ type~~ singularity            \cr
  =& {1 \over { 32 \pi^2 \alpha_1 \alpha_3}}\int_{S_C} d\sigma_{\nu \rho}
    \epsilon^{\mu \nu \rho \sigma} \hat{r}_\sigma (
    {{-2} \over {\mu^2 r^3}} - {1 \over r} K_2(\mu r) )
    + Coulomb~~ type~~ singularity                   \cr
  =& {1 \over { 32 \pi^2 \alpha_1 \alpha_3}}\int_{S_C} d\sigma_{\nu \rho}
    \epsilon^{\mu \nu \rho \sigma} \hat{r}_\sigma f(r)
    + Coulomb~~ type~~ singularity,
 }
  \eqn\eten
$$
where the Coulomb type singularity comes from the $<A_\mu A_\nu>$ propagator.
Since the function $f(r)$ behaves as follows
$f(r) \sim { 2 \over {\mu^2 r^3} }$ for $ r \gg { 1 \over \mu}$ and
$f(r) \sim  {\mu \over r}$ for $ r \ll { 1 \over \mu}$
the field strength  is smooth and concentrated to  a region
$ \sim { 1 \over \mu}$ around the  border $C$. The
string-like magnetic  flux present in  the topological $BF$  theory is
distributed over a  region with size $ \sim r \ll { 1 \over \mu}$ near the
border of
the open string world  sheet in (3+1)  dimension. Let the
border of  $ S_C $ be $C_1$ and the border of $ {S'}_C $ be $C_2$,
if $B_{\mu\nu}(x)$ is sufficiently far away  from
boundary we obtain the topological quantity $L$ where $L$ is the linking
number of
the two curves $C_1$ and $C_2$. If the curves are within a distance
$ \sim {1 \over \mu}$, the magnetic flux are overlapped, therefore
the statistical  interpretation for the  phase fails. By  adding the  
$F_{\mu \nu} F^{\mu \nu}$ and $H_{\mu \nu \rho}H^{\mu \nu \rho}$
terms to the action, the expectation value of Wilson surface
can be unambiguously defined; the spin factor is metric dependent but framing
independent. In the limit $ \mu \to \infty $ where gauge quanta become heavy,
the  long range interaction parts can be ignored, the phase factor leads to
the writhe.

Since in the pure theory without higher order terms the metric enter
the propagator via  gauge fixing, now we consider the gauge fixing
without the metric  dependence. In time like gauge, $A_0=0$ and $B_{0 \mu}=0$,
we obtain the spin factor,
$$
   \Phi^0 (S,C)
   = \oint_C dx_\mu \int_{ {\sigma'}_C} d\sigma_{\nu \rho}
   \epsilon_{\mu \nu \rho 0} \Theta ( x^0 - y^0 ) \delta^3 (x^i - y^i),
   \eqn\eeleven
$$
where $i = 1, 2, 3 $ and $\Theta$ is the step function.
Though $\Phi^0 (S,C)$ does not depend on the metric but it
is ill-defined ( contained $\delta^3 (0)$ ) and  should be regularized.
If we use the regularization
$\delta (x) \to  { 1 \over {\sqrt{2 \pi \epsilon}} } exp( -x^2 / \epsilon )$
as in Ref.\POLYAKOV , such regularization will give the writhe
and hence the metric dependence of  $ \Phi^0$ inevitable.

In discussion, so far we have analyzed the regularizations of the
expectation values of
Wilson surfaces. However it is required the more intimate discussion about
the gauge invariance of this theory, because the Wilson surface operator with
boundaries may not be invariant under the gauge transformation\efour\ with
respect to
two form $B_{\mu \nu} $. In order to deal with a problem of gauge invariance
more intimately, first, recall that in this paper we have introduced two kinds
of wilson surface(line) operators:
$$
  W_B = exp(i \int_S d \sigma^{\mu \nu} B_{\mu \nu} )
 \eqn \etwelve
$$
$$
  W_A = exp(i \oint_C dx^{\mu} A_{\mu} )
 \eqn \ethirteen
$$
As it was shown in Ref.\BLAU, the expectation value of either one of these equal
to 1 with respect to pure CS action. The non-trivial expectation values occur
from $<W_A W_B>$ where the expectation value is, of course, taken with
repect to the pure action\etwo.

In the fundamental string theory with the open bosonic sector, the antisymmetric
tensor gauge invariance must be act on the photon as well. The correct gauge
transformation
is as follows:
$$\eqalign
  {
  \delta B_{\mu \nu} = & \p_\mu \xi_\nu - \p_\nu \xi_\mu  \cr
  \delta A_\mu  =& -2\xi_\mu
  }
  \eqn\efourteen
$$
At string world sheet level this appear since a surface term for the variation
of $ B_{\mu \nu} $ must be cancelled by variation of the open string vector.
And there, of course, exists still the usual gauge invariance of vector $A_\mu$
independently.
We can, however, see that under this gauge transformation the pure action is not
invariant and hence some correct form of the pure action required. Fortunately
we can find the correct gauge invariant form of the action under the gauge
transformation\efourteen\ as the following modified form:
$$
  S_M = \int d^4 x \epsilon^{\mu \nu \rho \sigma} B_{\mu \nu} (\p_\rho A_\sigma
        + {1 \over 2} B_{\rho \sigma })
  \eqn\efifteen
$$
It is easy to see that the above action\efifteen\ is manifestly invariant
under the gauge
transformation\efourteen. By replacing $dA$ of $dA+B$ in form notation, it
is possible to
construct the modified action with higher order terms invariant under
the gauge transformation\efourteen. In spite of this change of symmetry, we
can still
show that the singular behavior of propagator for the system with gauge
symmetry\ethree\efour\ is identical to that of the system with gauge
symmetry\efifteen.
We think, however, as to the modified gauge symmetry the more detail
investigations will
be needed\LEE.

On the other hand, in closed string theory we can give the definition of
Wilson surface
for a closed surface $S$ and a loop $C$ as that of  Ref.\BLAU, however the
interaction
of Wilson line(surface) is invisible because the photon field $ A_\mu $ is not
appeared explicitly in the closed bosonic string effective theory\FRADKIN.

In conclusion, we have given the two definitions eq.(6) and eq.(5),(7) of the
expectation values of wilson surfaces operator through the different
regularizations
which are the ribbon splitting and the gauge invariant higher order term adding.
The regularization by adding gauge invariant higher order terms is manifestly
gauge invariant and the ribbon splitting is also gauge invariant,
so we think that the gauge invariance can not define the regularization
uniquely.

The difference between the two definitions eq.(6) and eq.(5)(7) can be resolved
by noting the fact that the limit $\epsilon \to 0 $ in eq.(6) is not smooth, 
since we have the relation $ W(writhe) = L(link) - T(torsion)$ for general
twisted ribbon.
The definition(6) is a topological invariant whose values depend on the framing
$\hat{n}$, while the definition(5)(7) by adding the higher oder terms lead to
the writhe whose values depend on the metric.
We think the existence of two definitions (6), (5)(7) originates in the
singular behavior
of propagator in the short distance, and the reason why the propagator is
singular is
due to the fact that the curve on which $A_\mu (x)$ field is defined must tend
to the border of the string world sheet on  which the antisymmetric $B_{\mu
\nu}$
field is defined.

The author is grateful to Jinho Cho and Hyeonjoon Shin for many helpful
discussions.
The author also thanks Won-Tae Kim for some remarks on regularizations.

\vfill\eject

\refout

\end